\documentclass[12pt]{article}
\usepackage{graphicx}
\usepackage{color}
\usepackage{cite}

\def \beq{\begin{equation}}
\def \eeq{\end{equation}}
\def\eqref#1{(\ref{#1})}
\def\bea{\begin{eqnarray}}
\def\eea{\end{eqnarray}}

\def\URLtilde{\lower0.2em\hbox{$\tilde{\phantom{a}}$}}
\def\mycomm#1{\hfill\break\strut\kern-3em{\color{red}\tt ====> #1
\color{black}}\hfill\break}

\newcount\timecount
\newcount\hours \newcount\minutes  \newcount\temp \newcount\pmhours
\hours = \time
\divide\hours by 60
\temp = \hours
\multiply\temp by 60
\minutes = \time
\advance\minutes by -\temp
\def\hour{\the\hours}
\def\minute{\ifnum\minutes<10 0\the\minutes
\else\the\minutes\fi}
\def\clock{
\ifnum\hours=0 12:\minute\ AM
\else\ifnum\hours<12 \hour:\minute\ AM
\else\ifnum\hours=12 12:\minute\ PM
\else\ifnum\hours>12
\pmhours=\hours
\advance\pmhours by -12
\the\pmhours:\minute\ PM
\fi
\fi
\fi
\fi
}

\def\monthname{\relax\ifcase\month 0/\or January\or February\or
March\or April\or May\or June\or July\or August\or September\or
October\or November\or December\else\number\month/\fi}

\def\bold#1{\setbox0=\hbox{$#1$}     \kern-.025em\copy0\kern-\wd0
\kern.05em\copy0\kern-\wd0
\kern-.025em\raise.0433em\box0 }

\begin{document}
\setcounter{footnote}{1}
\rightline{EFI 19-5}
\rightline{TAUP 3042/19}
\rightline{arXiv:1912.03204}
\vskip 0.7cm

\centerline{\large \bf Mass inequalities for baryons with heavy quarks}
\bigskip

\centerline{Marek Karliner$^a$\footnote{{\tt marek@proton.tau.ac.il}}
 and Jonathan L. Rosner$^b$\footnote{{\tt rosner@hep.uchicago.edu}}}
\medskip

\centerline{$^a$ {\it School of Physics and Astronomy}}
\centerline{\it Faculty of Exact Sciences}
\centerline{\it Tel Aviv University, Tel Aviv 69978, Israel}
\medskip

\centerline{$^b$ {\it Enrico Fermi Institute and Department of Physics}}
\centerline{\it University of Chicago, 5620 S. Ellis Avenue, Chicago, IL
60637, USA}
\bigskip
\strut

\begin{quote}
\begin{center}
ABSTRACT
\end{center}
Baryons with one or more heavy quarks have been shown, in the
context of a nonrelativistic description, to exhibit mass inequalities under
permutations of their quarks, when spin averages are taken.  These inequalities
sometimes are invalidated when spin-dependent forces are taken into account.
A notable instance is the inequality $2E(Mmm) > E(MMm) + E(mmm)$, where $m =
m_u = m_d$, satisfied for $M = m_b$ or $M = m_c$ but not for $M = m_s$,
unless care is taken to remove effects of spin-spin interactions.  Thus
in the quark-level analog of nuclear fusion, the reactions \,$\Lambda_b \Lambda_b
\to \Xi_{bb}N$\, and \,$\Lambda_c \Lambda_c \to \Xi_{cc}^{++}n$ \,are exothermic,
releasing respectively 138 and 12 MeV, while \,$\Lambda \Lambda \to \Xi N$\, is
endothermic, requiring an input of between 23 and 29 MeV.  Here we explore
such mass inequalities in the context of an approach, previously shown
to predict masses successfully, in which contributions consist of additive
constituent-quark masses, spin-spin interactions, and additional binding terms
for pairs each member of which is at least as heavy as a strange quark.
\end{quote}

\smallskip

\leftline{PACS codes: 14.20.Lq, 14.20.Mr, 12.40.Yx}
\bigskip

% \draft

% This is Section I
\section{INTRODUCTION \label{sec:intro}}

Quantum chromodynamics (QCD) predicts the existence of baryons containing
not only one, but two or three heavy quarks ($c$ or $b$).  The LHCb
Collaboration at CERN has discovered the first doubly-heavy baryon,
a\, $\Xi_{cc}^{++} = ccu$ \,state in the
decay modes $\Lambda_c K^- \pi^+ \pi^+$ \cite{Aaij:2017ueg} and $\Xi_c^+ \pi^+$
\cite{Aaij:2018gfl}, at a mass $M(\Xi_{cc}^{++}) = 3621.24 \pm 0.65 \pm 0.31$
MeV, very close to that predicted in Ref.\ \cite{Karliner:2014gca}.  (Hints of
a possible isospin partner $\Xi_{cc}^+$ decaying to $\Lambda_c K^- \pi^+$ have
also been found \cite{Aaij:2019jfq,KR2019}.)  In this approach, one adds up
constituent-quark masses and spin-spin interactions \cite{DeRujula:1975qlm,%
Lipkin:1978eh,Gasiorowicz:1981jz} and corrects for the additional binding in
any quark pair both of whose members are at least as heavy as a strange quark
\cite{Karliner:2014gca}.

Under some circumstances hadrons satisfy mass inequalities associated with
permutations of their quarks \cite{Bertlmann:1979zs,Richard:1983mu,Nussinov:%
1983hb,Weingarten:1983uj,Witten:1983ut,Lieb:1985aw,Martin:1986da,%
Anwar:2017toa}.  For example, under
some conditions one expects $2E(Mmm) > E(MMm) + E(mmm)$, where $E$ denotes the
mass, $m = m_u = m_d$, and $M$ is the mass of a heavy quark, to apply to
spin-averaged states [cf. Eq.~(3) in Ref.~\cite{Martin:1986da}]. This inequality
may be invalidated when spin-dependent forces are taken into account; it holds
for $M = m_b$ or $M = m_c$ but not for $M = m_s$.  Thus, in the quark-level
analog of nuclear fusion \cite{Karliner:2017elp}, the reactions $\Lambda_b
\Lambda_b \to \Xi_{bb}N$ and $\Lambda_c \Lambda_c\to\Xi_{cc}^{++}n$
are exothermic, releasing respectively 138 and 12 MeV, while $\Lambda \Lambda
\to \Xi N$ is endothermic, requiring energy input of between 23 and 29 MeV,
depending on which members of the $N$ and $\Xi$ doublets one uses.  

Here we give some examples of inequalities involving baryon masses in our
constituent-quark approach.  We outline in Sec.\ \ref{sec:rels} some of the
relations and their origin.  In Section \ref{sec:light} we treat light-quark
systems in which the strange quark plays the role of a heavy quark.  Baryons
with one and two heavy quarks ($c,b$) are described in Secs.\ \ref{sec:one}
and \ref{sec:two}, respectively, while Sec.\ \ref{sec:concl} concludes.

\section{Inequalities and their origin} \label{sec:rels}

A number of mass inequalities involving ground-state mesons and baryons were
noted by Nussinov \cite{Nussinov:1983hb}:  respectively,
\beq \label{eqn:neq}
m_{x \bar y} > \frac{1}{2}(m_{x \bar x} + m_{y \bar y})~,
\quad
m_{x y y} > \frac{1}{2}(m_{xxy} + m_{yyy})~.
\eeq
We shall motivate these relations in a simple case where quark masses enter
through their nonrelativistic kinetic energy, but they are much more general
(see many of the references quoted above, in particular \cite{Martin:1986da}).
Consider systems governed by the Hamiltonians
\beq
H_{ij} = \frac{\bold{p}^2}{2\mu_{ij}} + V(\bold{x})~,~~
\mu_{ij}\equiv\frac{m_1 m_2}{m_1+ m_2}~.
\eeq
Then, since
\beq
\frac{1}{\mu_{12}} = \frac{1}{m_1} + \frac{1}{m_2}~,
\eeq
we have $H_{12} = \frac{1}{2} (H_{11} + H_{22})$.  The ground-state energy
in the 12 system is
\beq
E_{12} = {\rm Min}_\psi \langle \psi_{12}|H_{12}|\psi_{12} \rangle = \langle
\psi_{12}|\frac{1}{2} \left(H_{11} + H_{22} \right)|\psi_{12} \rangle~.
\eeq
Now
\beq
\langle \psi_{12}|H_{11}|\psi_{12} \rangle > E_1~;
\quad
\langle \psi_{12}|H_{22}|\psi_{12} \rangle > E_2~,
\eeq
since $\psi_{12}$ is not the ground state (assumed non-degenerate) of $H_{11}$
or $H_{22}$.  Hence we have the result $E_{12} > (E_1+E_2)/2$, implying
Eq.\ (\ref{eqn:neq}).

When spin-spin interactions are taken into account, the energy shift due
to the interaction of quarks $i$ and $j$ may be written
\beq \label{eqn:hfs}
\Delta E_{ij,{\rm HFs}}=b\langle \sigma_i \cdot \sigma_j \rangle /(m_i m_j)~.
\eeq
Inclusion of such terms in the Hamiltonian in general spoils the relation
\hbox{$H_{12} = (H_{11}+H_{22})/2$.}

\section{Baryons with light quarks} \label{sec:light}

If we adopt the semi-empirical model of hadron masses \cite{Karliner:2014gca,%
Karliner:2019lau} in which the constituent $u,d,s$ quarks have constituent
masses of several hundred MeV, the only corrections to the sum of quark masses
are the sum of hyperfine terms (\ref{eqn:hfs}) and terms $B(ss)$ expressing the
stronger binding of one strange quark with another.  Here the tilde stands for
a mass without hyperfine interaction terms.  The Gell-Mann--Okubo relation
 \hbox{${1\over2}[M(N){+}M(\Xi)]{=}{1\over4}[3M(\Lambda){+}M(\Sigma)]$,}
derived under the assumption that SU(3) breaking is linear in hypercharge,
then becomes an inequality
\beq
\frac{1}{2} \left[ \tilde M(N) + \tilde M(\Xi) \right] = 2 m_q + m_s - B(ss)/2
 < \frac{1}{4} \left[3\tilde M(\Lambda) + \tilde M(\Sigma) \right] = 2 m_q
 + m_s~,
\eeq
and the second inequality in Eq.\ (\ref{eqn:neq}) is satisfied.  (It turns out
that the experimental values of octet baryon masses \cite{Tanabashi:2018oca}
also satisfy this relation, with $[M(N) + M(\Xi)]/2 = 1128.6$ MeV and
$[3M(\Lambda) + M(\Sigma)]/4 = 1135.0$ MeV.)

When spin-spin interactions [Eq.\ (\ref{eqn:hfs})] are taken into account
as in Ref.\ \cite{Karliner:2014gca}, one has
\beq \label{eqn:diff}
\frac{1}{2} \left[ \tilde M(N) + \tilde M(\Xi) \right] - \frac{1}{4} \left[
3\tilde M(\Lambda) + \tilde M(\Sigma) \right] = -\frac{B(ss)}{2} 
+ b \frac{(m_s-m_q)^2}{2m_s^2 m_q^2}~,
\eeq
where $m_q$ stands for the isospin-averaged mass of $u$ and $d$ quarks.
A fit to masses of light-quark baryons \cite{Karliner:2019lau} gives $b/m_q^2 =
50.0$ MeV, $m_q = 362.1$ MeV, $m_s = 543.9$ MeV, $B(ss) = 9.23$ MeV, so the
right-hand side of Eq.\ (\ref{eqn:diff}) is $-4.6$ MeV + 2.8 MeV, with the
spin-spin term working against the predicted inequality but not enough to
counteract it. 

\section{Baryons with one heavy quark} \label{sec:one}

The masses of ground state baryons $cqq$, $cqs$, and $css$ containing a single
charmed quark satisfy the inequality
\beq \label{eqn:cbar}
[\tilde M(\Lambda_c) + \tilde M(\Omega_c)]/2 < \tilde M(\Xi_c)~,
\eeq
 thanks to the binding correction $B(ss) = 9.2$ MeV \cite{Karliner:2019lau}:%
\footnote{A term $B(cs) =
35.0$ MeV \cite{Karliner:2014gca} is common to both sides.}
\bea
[\tilde M(\Lambda_c) + \tilde M(\Omega_c)]/2 & = & m_q+m_s+ m_c -B(cs)-B(ss)/2
\nonumber \\ < \tilde M(\Xi_c) & = & m_q + m_s + m_c - B(cs)~.
\eea
The experimental values \cite{Tanabashi:2018oca} do not satisfy the inequality:
\beq
\frac{1}{2}[M(\Lambda_c) + M(\Omega_c)] = 2490.8~{\rm MeV} > M(\Xi_c)
= 2468.9~{\rm MeV}~,
\eeq
indicating a significant contribution from spin-spin interactions.  (Here we
have used the isospin-averaged $\Xi_c$ mass.)

The corresponding relation for states containing a $b$ quark is
\bea
[\tilde M(\Lambda_b) + \tilde M(\Omega_b)]/2 & = & m_q+m_s+ m_b -B(bs)-B(ss)/2
\nonumber \\ < \tilde M(\Xi_b) & = & m_q + m_s + m_b - B(bs)~,
\eea
with $B(bs) = 41.8$ MeV \cite{Karliner:2014gca} common to both sides.  Again,
the term $B(ss)$ turns the equality into an inequality.  The experimental
values, again thanks to a significant spin-spin contribution, satisfy the
opposite inequality:
\beq
\frac{1}{2}[M(\Lambda_b) + M(\Omega_b)] = 5832.9~{\rm MeV} > M(\Xi_b)
= 5791.3~{\rm MeV}~.
\eeq

\section{Baryons with two or more heavy quarks} \label{sec:two}

In this section we discuss inequalities between binding energies.
In our convention the binding energies are positive, but they contribute
to the hadron mass with a minus sign, so the direction of the
inequalities between binding energies is flipped vs. inequalities between 
hadron masses.

We have seen that the second of Eqs.\ (\ref{eqn:neq}) holds for baryons
with zero or one heavy ($c,b$) quarks.  Inequalities also hold in our
semi-empirical approach when two quarks are heavy.  For
example \cite{Karliner:2014gca,Karliner:2019lau}, $B(ss) = 9.2$ MeV,
$B(cs) = 35.0$ MeV, $B(cc) = 129.0$ MeV, so
\beq
B(cs) < [B(ss) + B(cc)]/2~.
\eeq
Similarly $B(bs) = 41.8$ MeV, $B(bb) = 281.4$ MeV, so
\beq
B(bs) < [B(ss) + B(bb)]2~.
\eeq
It was also found \cite{Karliner:2014gca} that $B(bc) = 170.8$ MeV, so
\beq
 B(bc) = 170.8~{\rm MeV} < [B(cc) + B(bb)]/2 = 205.2~{\rm MeV}~.
\eeq
A related inequality for the spin-weighted averages $\bar M$ of heavy $c \bar
c$, $b \bar b$, and $b \bar c$ ground-state mesons,
\beq
[\bar M(c \bar c) + \bar M(b \bar b)]2 < \bar M(b \bar c)~,
\eeq
  is comfortably satisfied: The left-hand side is $[3068.7 + 9444.9]/2=6256.8$
MeV, while the right-hand side is likely \cite{Eichten:2019gig} to be tens of
 MeV above the mass $M(B_c) = 6274.9 \pm 0.8$ MeV \cite{Tanabashi:2018oca} of
the spin-zero pseudoscalar $b \bar c$ ground state.

\section{Conclusions} \label{sec:concl}

A semi-empirical method \cite{Karliner:2014gca} successfully determines hadron
masses, including the mass of the first doubly charmed baryon.  This method
sums constituent-quark masses, quark-quark hyperfine interactions, and terms
$B(qq')$ expressing the binding of quarks both of which are at least as heavy
as the strange quark.  These binding terms are seen to satisfy inequalities
$B(qq') < [B(qq) + B(q'q')]/2$, with the consequence that when hyperfine
contributions are removed, baryons satisfy the inequality $m_{x y y} >
\frac{1}{2}(m_{xxy} + m_{yyy})$ \cite{Bertlmann:1979zs,Richard:1983mu,%
Nussinov:1983hb,Weingarten:1983uj,Witten:1983ut,Lieb:1985aw,Martin:1986da,%
Anwar:2017toa}.
This constitutes a useful consistency check of the semi-empirical method, and
enables rough estimates, independent of potential models, of unseen hadron
masses.

\section*{ACKNOWLEDGMENTS}
The research of M.K. was supported in part by NSFC-ISF grant No.~0603219411.
We thank J.-M. Richard for a helpful discussion.

\end{document}